\documentstyle[10pt,epsf,epsfig,hangcaption,xspace,amssymb,amsfonts,amsmath,amsthm,cite,
dp_delphititle]{dp_delphi}
\makeindex
\def\DpPaperGroup{EP}
\def\DpPaperRef{2000-009}
\def\DpDate{18 January 2000}
\def\DpAuthors{DELPHI Collaboration}
\def\DpSubmit{(Phys. Letters B475 (2000)429)}
\def\DpTitle{{Inclusive \boldmath $\Sigma^-$ and $\lft$ 
           production \\ in hadronic Z decays}}
\def\DpComment{ }
\def\DpEMail{} 

\newcommand{\s}{{\Sigma}\xspace}
\newcommand{\spm}{{\Sigma^{\pm}}\xspace}
\renewcommand{\sp}{{\Sigma^{+}}\xspace}
\newcommand{\sm}{{\Sigma^{-}}\xspace}

\newcommand{\sdec}{{\Sigma\rightarrow\mbox{n}\pi}\xspace}
\newcommand{\spmdec}{{\Sigma^{\pm}\rightarrow\mbox{n}\pi^{\pm}}\xspace}
\newcommand{\smdec}{{\Sigma^-\rightarrow\mbox{n}\pi^-}\xspace}

\newcommand{\lft}{{\Lambda(1520)}\xspace}
\newcommand{\ldec}{{\Lambda(1520)\rightarrow{\rm pK^-}}\xspace}
\newcommand{\gev}{{\ifmmode \mbox{Ge\kern-0.2exV}
\else Ge\kern-0.2exV\nolinebreak\fi}}
\newcommand{\mev}{{\ifmmode \mbox{Me\kern-0.2exV}
\else Me\kern-0.2exV\nolinebreak\fi}}
\newcommand{\degr}{$^\circ$\xspace}
\newcommand{\jst}{\textsc{Jetset 7.3}\xspace}
\newcommand{\jsf}{\textsc{Jetset 7.4}\xspace}
\newcommand{\herwig}{\textsc{Herwig 5.9}\xspace}
\newcommand{\jetset}{\textsc{Jetset}\xspace}
\newcommand{\delphi}{\textsc{Delphi}\xspace}
\newcommand{\alephc}{\textsc{Aleph}\xspace}
\newcommand{\opal}{\textsc{Opal}\xspace}
\newcommand{\lthree}{\textsc{L3}\xspace}
\newcommand{\lep}{\textsc{Lep}\xspace}
\newcommand{\delsim}{\textsc{Delsim}\xspace}
\begin{document}
\makeatletter
\newcount\@tempcntc
\def\@citex[#1]#2{\if@filesw\immediate\write\@auxout{\string\citation{#2}}\fi
  \@tempcnta\z@\@tempcntb\m@ne\def\@citea{}\@cite{\@for\@citeb:=#2\do
    {\@ifundefined
       {b@\@citeb}{\@citeo\@tempcntb\m@ne\@citea\def\@citea{,}{\bf ?}\@warning
       {Citation `\@citeb' on page \thepage \space undefined}}%
    {\setbox\z@\hbox{\global\@tempcntc0\csname b@\@citeb\endcsname\relax}%
     \ifnum\@tempcntc=\z@ \@citeo\@tempcntb\m@ne
       \@citea\def\@citea{,}\hbox{\csname b@\@citeb\endcsname}%
     \else
      \advance\@tempcntb\@ne
      \ifnum\@tempcntb=\@tempcntc
      \else\advance\@tempcntb\m@ne\@citeo
      \@tempcnta\@tempcntc\@tempcntb\@tempcntc\fi\fi}}\@citeo}{#1}}
\def\@citeo{\ifnum\@tempcnta>\@tempcntb\else\@citea\def\@citea{,}%
  \ifnum\@tempcnta=\@tempcntb\the\@tempcnta\else
   {\advance\@tempcnta\@ne\ifnum\@tempcnta=\@tempcntb \else \def\@citea{--}\fi
    \advance\@tempcnta\m@ne\the\@tempcnta\@citea\the\@tempcntb}\fi\fi}
 
\makeatother
\begin{titlepage}
\pagenumbering{roman}
\CERNpreprint{\DpPaperGroup}{\DpPaperRef} 
\date{{\small\DpDate}} 
\title{\DpTitle} 
\address{\DpAuthors} 
\begin{shortabs} 
\noindent
%
\noindent

Production of $\sm$ and $\lft$ in hadronic Z decays has been measured  
using the
\delphi detector at \lep. The $\sm$ is directly reconstructed as a
charged track in the \delphi microvertex detector and is identified by its
$\sdec$ decay leading to a kink between the $\sm$ and $\pi$-track. 
The reconstruction of the $\lft$ resonance relies strongly on the
particle identification capabilities of the barrel Ring Imaging Cherenkov
detector and on the ionisation loss measurement of the TPC.
Inclusive production spectra are measured for both particles. 
The production
rates are measured to be
\begin{eqnarray*}
\langle N_{\sm}/N_{\rm Z}^{\rm had} \rangle &=& 0.081 \pm 0.002 \pm  0.010~,  \\
\langle N_{\lft}/N_{\rm Z}^{\rm had} \rangle &=& 0.029 \pm 0.005 \pm  0.005~.
\end{eqnarray*}
The production rate of the $\lft$ suggests
that a large fraction of the stable baryons descend from orbitally excited 
baryonic states.
It is shown that the baryon production rates in Z decays follow
a universal phenomenological law related to isospin, strangeness and
mass of the particles.

\end{shortabs}
\vfill
\begin{center}
\DpSubmit \ \\ 
\DpComment \ \\
\DpEMail \ \\
\end{center}
\vfill
\clearpage
\headsep 10.0pt
\addtolength{\textheight}{10mm}
\addtolength{\footskip}{-5mm}
\begingroup
%
\newcommand{\DpName}[2]{\hbox{#1$^{\ref{#2}}$},\hfill}
\newcommand{\DpNameTwo}[3]{\hbox{#1$^{\ref{#2},\ref{#3}}$},\hfill}
\newcommand{\DpNameThree}[4]{\hbox{#1$^{\ref{#2},\ref{#3},\ref{#4}}$},\hfill}
\newskip\Bigfill \Bigfill = 0pt plus 1000fill
\newcommand{\DpNameLast}[2]{\hbox{#1$^{\ref{#2}}$}\hspace{\Bigfill}}
%
\footnotesize
\noindent
\DpName{P.Abreu}{LIP}
\DpName{W.Adam}{VIENNA}
\DpName{T.Adye}{RAL}
\DpName{P.Adzic}{DEMOKRITOS}
\DpName{I.Ajinenko}{SERPUKHOV}
\DpName{Z.Albrecht}{KARLSRUHE}
\DpName{T.Alderweireld}{AIM}
\DpName{G.D.Alekseev}{JINR}
\DpName{R.Alemany}{VALENCIA}
\DpName{T.Allmendinger}{KARLSRUHE}
% ...
\normalsize
\endgroup
\titlefoot{Department of Physics and Astronomy, Iowa State
     University, Ames IA 50011-3160, USA
    \label{AMES}}
\titlefoot{Physics Department, Univ. Instelling Antwerpen,
     Universiteitsplein 1, B-2610 Antwerpen, Belgium \\
     \indent~~and IIHE, ULB-VUB,
     Pleinlaan 2, B-1050 Brussels, Belgium \\
     \indent~~and Facult\'e des Sciences,
     Univ. de l'Etat Mons, Av. Maistriau 19, B-7000 Mons, Belgium
    \label{AIM}}
\titlefoot{Physics Laboratory, University of Athens, Solonos Str.
     104, GR-10680 Athens, Greece
    \label{ATHENS}}
\titlefoot{Department of Physics, University of Bergen,
     All\'egaten 55, NO-5007 Bergen, Norway
    \label{BERGEN}}
\titlefoot{Dipartimento di Fisica, Universit\`a di Bologna and INFN,
     Via Irnerio 46, IT-40126 Bologna, Italy
    \label{BOLOGNA}}
\titlefoot{Centro Brasileiro de Pesquisas F\'{\i}sicas, rua Xavier Sigaud 150,
     BR-22290 Rio de Janeiro, Brazil \\
     \indent~~and Depto. de F\'{\i}sica, Pont. Univ. Cat\'olica,
     C.P. 38071 BR-22453 Rio de Janeiro, Brazil \\
     \indent~~and Inst. de F\'{\i}sica, Univ. Estadual do Rio de Janeiro,
     rua S\~{a}o Francisco Xavier 524, Rio de Janeiro, Brazil
    \label{BRASIL}}
\titlefoot{Comenius University, Faculty of Mathematics and Physics,
     Mlynska Dolina, SK-84215 Bratislava, Slovakia
    \label{BRATISLAVA}}
\titlefoot{Coll\`ege de France, Lab. de Physique Corpusculaire, IN2P3-CNRS,
     FR-75231 Paris Cedex 05, France
    \label{CDF}}
\titlefoot{CERN, CH-1211 Geneva 23, Switzerland
    \label{CERN}}
\titlefoot{Institut de Recherches Subatomiques, IN2P3 - CNRS/ULP - BP20,
     FR-67037 Strasbourg Cedex, France
    \label{CRN}}
\titlefoot{Now at DESY-Zeuthen, Platanenallee 6, D-15735 Zeuthen, Germany
    \label{DESY}}
\titlefoot{Institute of Nuclear Physics, N.C.S.R. Demokritos,
     P.O. Box 60228, GR-15310 Athens, Greece
    \label{DEMOKRITOS}}
\titlefoot{FZU, Inst. of Phys. of the C.A.S. High Energy Physics Division,
     Na Slovance 2, CZ-180 40, Praha 8, Czech Republic
    \label{FZU}}
\titlefoot{Dipartimento di Fisica, Universit\`a di Genova and INFN,
     Via Dodecaneso 33, IT-16146 Genova, Italy
    \label{GENOVA}}
\titlefoot{Institut des Sciences Nucl\'eaires, IN2P3-CNRS, Universit\'e
     de Grenoble 1, FR-38026 Grenoble Cedex, France
    \label{GRENOBLE}}
\titlefoot{Helsinki Institute of Physics, HIP,
     P.O. Box 9, FI-00014 Helsinki, Finland
    \label{HELSINKI}}
\titlefoot{Joint Institute for Nuclear Research, Dubna, Head Post
     Office, P.O. Box 79, RU-101 000 Moscow, Russian Federation
    \label{JINR}}
\titlefoot{Institut f\"ur Experimentelle Kernphysik,
     Universit\"at Karlsruhe, Postfach 6980, DE-76128 Karlsruhe,
     Germany
    \label{KARLSRUHE}}
\titlefoot{Institute of Nuclear Physics and University of Mining and Metalurgy,
     Ul. Kawiory 26a, PL-30055 Krakow, Poland
    \label{KRAKOW}}
\titlefoot{Universit\'e de Paris-Sud, Lab. de l'Acc\'el\'erateur
     Lin\'eaire, IN2P3-CNRS, B\^{a}t. 200, FR-91405 Orsay Cedex, France
    \label{LAL}}
\titlefoot{School of Physics and Chemistry, University of Lancaster,
     Lancaster LA1 4YB, UK
    \label{LANCASTER}}
\titlefoot{LIP, IST, FCUL - Av. Elias Garcia, 14-$1^{o}$,
     PT-1000 Lisboa Codex, Portugal
    \label{LIP}}
\titlefoot{Department of Physics, University of Liverpool, P.O.
     Box 147, Liverpool L69 3BX, UK
    \label{LIVERPOOL}}
\titlefoot{LPNHE, IN2P3-CNRS, Univ.~Paris VI et VII, Tour 33 (RdC),
     4 place Jussieu, FR-75252 Paris Cedex 05, France
    \label{LPNHE}}
\titlefoot{Department of Physics, University of Lund,
     S\"olvegatan 14, SE-223 63 Lund, Sweden
    \label{LUND}}
\titlefoot{Universit\'e Claude Bernard de Lyon, IPNL, IN2P3-CNRS,
     FR-69622 Villeurbanne Cedex, France
    \label{LYON}}
\titlefoot{Univ. d'Aix - Marseille II - CPP, IN2P3-CNRS,
     FR-13288 Marseille Cedex 09, France
    \label{MARSEILLE}}
\titlefoot{Dipartimento di Fisica, Universit\`a di Milano and INFN-MILANO,
     Via Celoria 16, IT-20133 Milan, Italy
    \label{MILANO}}
\titlefoot{Dipartimento di Fisica, Univ. di Milano-Bicocca and
     INFN-MILANO, Piazza delle Scienze 2, IT-20126 Milan, Italy
    \label{MILANO2}}
\titlefoot{Niels Bohr Institute, Blegdamsvej 17,
     DK-2100 Copenhagen {\O}, Denmark
    \label{NBI}}
\titlefoot{IPNP of MFF, Charles Univ., Areal MFF,
     V Holesovickach 2, CZ-180 00, Praha 8, Czech Republic
    \label{NC}}
\titlefoot{NIKHEF, Postbus 41882, NL-1009 DB
     Amsterdam, The Netherlands
    \label{NIKHEF}}
\titlefoot{National Technical University, Physics Department,
     Zografou Campus, GR-15773 Athens, Greece
    \label{NTU-ATHENS}}
\titlefoot{Physics Department, University of Oslo, Blindern,
     NO-1000 Oslo 3, Norway
    \label{OSLO}}
\titlefoot{Dpto. Fisica, Univ. Oviedo, Avda. Calvo Sotelo
     s/n, ES-33007 Oviedo, Spain
    \label{OVIEDO}}
\titlefoot{Department of Physics, University of Oxford,
     Keble Road, Oxford OX1 3RH, UK
    \label{OXFORD}}
\titlefoot{Dipartimento di Fisica, Universit\`a di Padova and
     INFN, Via Marzolo 8, IT-35131 Padua, Italy
    \label{PADOVA}}
\titlefoot{Rutherford Appleton Laboratory, Chilton, Didcot
     OX11 OQX, UK
    \label{RAL}}
\titlefoot{Dipartimento di Fisica, Universit\`a di Roma II and
     INFN, Tor Vergata, IT-00173 Rome, Italy
    \label{ROMA2}}
\titlefoot{Dipartimento di Fisica, Universit\`a di Roma III and
     INFN, Via della Vasca Navale 84, IT-00146 Rome, Italy
    \label{ROMA3}}
\titlefoot{DAPNIA/Service de Physique des Particules,
     CEA-Saclay, FR-91191 Gif-sur-Yvette Cedex, France
    \label{SACLAY}}
\titlefoot{Instituto de Fisica de Cantabria (CSIC-UC), Avda.
     los Castros s/n, ES-39006 Santander, Spain
    \label{SANTANDER}}
\titlefoot{Dipartimento di Fisica, Universit\`a degli Studi di Roma
     La Sapienza, Piazzale Aldo Moro 2, IT-00185 Rome, Italy
    \label{SAPIENZA}}
\titlefoot{Inst. for High Energy Physics, Serpukov
     P.O. Box 35, Protvino, (Moscow Region), Russian Federation
    \label{SERPUKHOV}}
\titlefoot{J. Stefan Institute, Jamova 39, SI-1000 Ljubljana, Slovenia
     and Laboratory for Astroparticle Physics,\\
     \indent~~Nova Gorica Polytechnic, Kostanjeviska 16a, SI-5000 Nova Gorica, Slovenia, \\
     \indent~~and Department of Physics, University of Ljubljana,
     SI-1000 Ljubljana, Slovenia
    \label{SLOVENIJA}}
\titlefoot{Fysikum, Stockholm University,
     Box 6730, SE-113 85 Stockholm, Sweden
    \label{STOCKHOLM}}
\titlefoot{Dipartimento di Fisica Sperimentale, Universit\`a di
     Torino and INFN, Via P. Giuria 1, IT-10125 Turin, Italy
    \label{TORINO}}
\titlefoot{Dipartimento di Fisica, Universit\`a di Trieste and
     INFN, Via A. Valerio 2, IT-34127 Trieste, Italy \\
     \indent~~and Istituto di Fisica, Universit\`a di Udine,
     IT-33100 Udine, Italy
    \label{TU}}
\titlefoot{Univ. Federal do Rio de Janeiro, C.P. 68528
     Cidade Univ., Ilha do Fund\~ao
     BR-21945-970 Rio de Janeiro, Brazil
    \label{UFRJ}}
\titlefoot{Department of Radiation Sciences, University of
     Uppsala, P.O. Box 535, SE-751 21 Uppsala, Sweden
    \label{UPPSALA}}
\titlefoot{IFIC, Valencia-CSIC, and D.F.A.M.N., U. de Valencia,
     Avda. Dr. Moliner 50, ES-46100 Burjassot (Valencia), Spain
    \label{VALENCIA}}
\titlefoot{Institut f\"ur Hochenergiephysik, \"Osterr. Akad.
     d. Wissensch., Nikolsdorfergasse 18, AT-1050 Vienna, Austria
    \label{VIENNA}}
\titlefoot{Inst. Nuclear Studies and University of Warsaw, Ul.
     Hoza 69, PL-00681 Warsaw, Poland
    \label{WARSZAWA}}
\titlefoot{Fachbereich Physik, University of Wuppertal, Postfach
     100 127, DE-42097 Wuppertal, Germany
    \label{WUPPERTAL}}
\addtolength{\textheight}{-10mm}
\addtolength{\footskip}{5mm}
\clearpage
\headsep 30.0pt
\end{titlepage}
%
\pagenumbering{arabic} 
\setcounter{footnote}{0} %
\large
%
\section{Introduction}\label{s:intro}
The study of baryon production
provides an important tool to test models of the
fragmentation process \cite{lotsofreferencestobeput}. 
Beyond the cluster fragmentation model \cite{herwig}, and the 
string model\cite{jetset} which employs many
parameters to describe baryon fragmentation,
thermodynamical \cite{beccatini} and phenomenological models
\cite{chliapnikov,pei,chliap2}
have appeared recently which successfully describe
the overall particle production rates in
high energy interactions with very few parameters.

At \lep it has been shown that a large fraction of the observed stable
mesons stem from decays of scalar and tensor mesons with angular momentum.
For baryons this is not the case, as baryon resonances, especially those
with orbital excitations, typically have a large decay width and complicated
decay modes. Hence these states are difficult to access experimentally in a
multihadronic environment. In any case, it is still a question of
basic importance as to how far
baryon production leads to excited baryonic states.

So far the only orbitally excited baryonic state measured in $e^+e^-$
annihilation is the $\lft$ \cite{argus,opal_res_paper}.
This paper provides further data on $\lft$ and $\sm$
production\footnote{The antiparticles are always implicitly included.}
in hadronic Z decays.
For $\sm$ production at the Z so far only two measurements are
available~\cite{sigmaolddelphi,opalsigma}, one of them being sensitive
only to the sum of $\sm$ and $\sp$ states \cite{sigmaolddelphi}.

The data used throughout this paper were collected by the \delphi detector 
in 1994 and 1995.
In these data taking periods the \delphi microVertex (VD) and Ring Imaging
Cherenkov (RICH) detectors were optimally set up and functioning
for
the analyses presented.

This paper is organised as follows. Section~(\ref{s:det_ana}) gives a brief
overview on the detector, experimental procedures used to select tracks and
hadronic events as well as on the specific experimental procedures and
corrections used for
$\sm$ and $\lft$ reconstruction.
Section~(\ref{s:results}) presents
the results for $\sm$ and $\lft$ production and the corresponding systematic errors.
These results are compared to the expectation of 
fragmentation models and a general phenomenological law of baryon
production in Z decays is deduced. Finally we conclude in
Section~(\ref{s:conclusion}).

\section{The Experimental Procedure and Event Selection \label{s:det_ana}}

The \delphi detector is described in detail in~\cite{detector}.
The present analysis relies on information provided by
the central tracking detectors and the barrel RICH:
\begin{itemize}
\item The {\bf microVertex Detector} (VD) consists of three layers of silicon strip
      detectors at radii of 6.3, 9.0 and 10.9~cm.
      $R\phi$ coordinates\footnote{In the standard DELPHI coordinate system,
        the $z$ axis is along the electron direction, the $x$ axis points
        towards the centre of LEP, and the $y$ axis points upwards. The polar
        angle to the $z$ axis is denoted by $\theta$, and the azimuthal angle
        around the $z$ axis by $\phi$; the radial coordinate is $R=\sqrt{x^2+y^2}$.}
      in the plane perpendicular to the beam
      are measured in all three layers. The first and third layers also
      provide $z$ information.
      The polar angle ($\theta$) 
      coverage for a
      particle passing all three layers is from 44\degr to 136\degr.
      The single point resolution has been estimated from real data to be
      about 8 $\mu$m in $R\phi$ and (for charged particles crossing
      perpendicular to the module) about 9 $\mu$m in $z$.
\item The {\bf Inner Detector} (ID) consists of an inner drift chamber with
      jet chamber geometry and 5 cylindrical MWPC (in 1995 straw tube) layers. 
      The jet chamber,
      between 12 and 23~cm in $R$ and 23\degr and 157\degr
      (15\degr-165\degr for 1995) in 
      $\theta$, consists of 24 azimuthal sectors, each providing up to
      24 $R\phi$ points.
\item The {\bf Time Projection Chamber} (TPC) is the main tracking device of
      \delphi. It provides up to 16 space points per particle trajectory
      for radii between 40 and 110~cm. The precision on the track
      elements is about 150 $\mu$m in $R\phi$ and about 600 $\mu$m in $z$.
      A measurement of the energy loss $dE/dx$ of a track is provided with 
      a resolution of about 6.5\%.
\item The {\bf Outer Detector} (OD) is a 4.7~m long set of 5 layers of drift tubes
      situated at 2~m radius to the beam which provides precise spatial information
      in $R\phi$.
\item The {\bf Barrel Ring Imaging Cherenkov Counter} (BRICH) is 
      the main \delphi detector devoted to charged particle identification.
      It is subdivided into two halves ($ z \gtrless 0$) and provides 
      particle identification using Cherenkov radiation produced in a
      liquid or a gas radiator. This radiation, after appropriate focusing, is
      transformed into photoelectrons in a TPC-like drift structure and the
      Cherenkov angles of the track in both media are determined.
      The BRICH detector provides particle identification in the momentum range
      $0.7$ to $45$ GeV/$c$. 
\end{itemize}
An event was selected as a multihadronic event if the following
requirements were satisfied:
\begin{itemize}
\item There were at least five well measured charged particles in the event, 
      each with
      momentum larger than 300~MeV/$c$ (400~MeV/$c$ for the $\sm$ analysis)
      and a polar angle in the range $20\mbox{\degr}<\theta<160\mbox{\degr}$.
\item The total reconstructed energy of these
      charged tracks had to be larger than 12\% of the 
      centre-of-mass energy.
\item The total energy of the charged particles in each detector hemisphere
      (defined by the plane perpendicular to the beam axis)
      had to exceed 3\% of the centre-of-mass energy.
\item The tracking devices and, in the case of the $\lft$ analysis also the BRICH,
      were fully operational.
\end{itemize}

After these cuts, about 1.3 million events remained for the 
1994 period and 0.6 million events for the 1995 run around the Z$^0$ pole.
The $\lft$ analysis is based on both years data, the $\sm$ analysis only on
the 1994 data.

To study the influence of cuts, inefficiencies 
and resolution as well as particle
re-interactions in the detector, a large set of simulated
$\mbox{Z} \rightarrow \mbox{q} \bar{\mbox{q}}$ events has been used.
This simulated sample has been generated using 
the \jst model \cite{jetset} with the parton shower option.
The model parameters were taken 
from earlier QCD studies \cite{deltune}.
The initial event simulation was 
followed by a detailed detector simulation \cite{delsim}.
For the $\lft$ study a specific set of 10000 of 
these events has also been produced with
at least one $\ldec$ per event.

\subsection{\boldmath{$\spm$} Reconstruction}

The charged $\s$ hyperons decay through the weak interaction according to
\begin{eqnarray}
  \label{sigmaprop}
  \sp & \to \mbox{p}\pi^0 & (\approx 52\%) , \nonumber \\
      & \to \mbox{n}\pi^+ & (\approx 48\%) , \\
  \sm & \to \mbox{n}\pi^- & (\approx 100\%) \nonumber.
\end{eqnarray}
The reconstruction of the decay $\spmdec$ is based on
the large flight length ($c\tau$ = 4.43~cm for $\sm$ and $c\tau$ = 2.4~cm for $\sp$).
It allows a determination of the track parameters for the $\spm$,
if there are at least three hits in the microvertex detector.
The decay $\spmdec$ is then reconstructed by finding the
kink between the $\spm$ and the pion which is normally well measured in the
other tracking chambers, especially the TPC.
Thus, the detection of the neutron is not necessary.
The tools needed for the analysis are described in more detail below.

\subsubsection{Track reconstruction with the Microvertex Detector}

The geometry of the 1994 \delphi microvertex detector allows the determination
of the track parameters of a charged particle in two or three dimensions using 
information from this detector alone. These tracks, called 
`VD tracks' in the following, can arise from:
\begin{itemize}
\item Low-momentum charged particles 
      ($p \lesssim$~50~MeV/$c$) not reaching the TPC
      due to the bending in the magnetic field.
\item Tracks at the borders of the TPC modules not efficiently
      reconstructed by the standard tracking algorithm.
\item Charged particles interacting with the detector material outside the VD.
\item Decays such as $X^{\pm} \to Y^{\pm}$ + neutral particles 
      (e.g. $\spmdec$,
            $\Xi^- \to \Lambda \pi^-$,
            $\mbox{K}^{\pm}/\pi^{\pm} \to \mu^{\pm} \nu_{\mu}$).
\end{itemize}
To reconstruct these tracks, the microvertex tracking algorithm initially
looks for triplets of $R\phi$ hits in the VD (there must be at least one hit
in each layer), not associated to tracks
reconstructed by the standard tracking algorithm.
This allows a determination of the track parameters in the $R\phi$ plane.
A second step searches
for unassociated $z$ hits in the modules containing the $R\phi$ hits.
If there are at least two $z$ hits for a $R\phi$ triplet, the
full set of parameters is given for that track. To improve the
momentum resolution and to remove products of hadronic interactions in
the beampipe, the results from the VD tracking were refitted forcing the track
to originate from the primary vertex. The $\chi^2$ probability of this
fit had to be greater than 0.1\%.

\subsubsection{Efficiency correction procedure}\label{s:effcorr}

To be as independent as possible from the detector simulation, especially
from the modelling of the VD, the efficiency for reconstructing a VD track
has been deduced directly from the data as follows.

\begin{figure}[t]
\begin{minipage}[t]{0.6\textwidth}  
\begin{flushleft}
\epsfig{bb=30 51 561 404,width=\textwidth,file=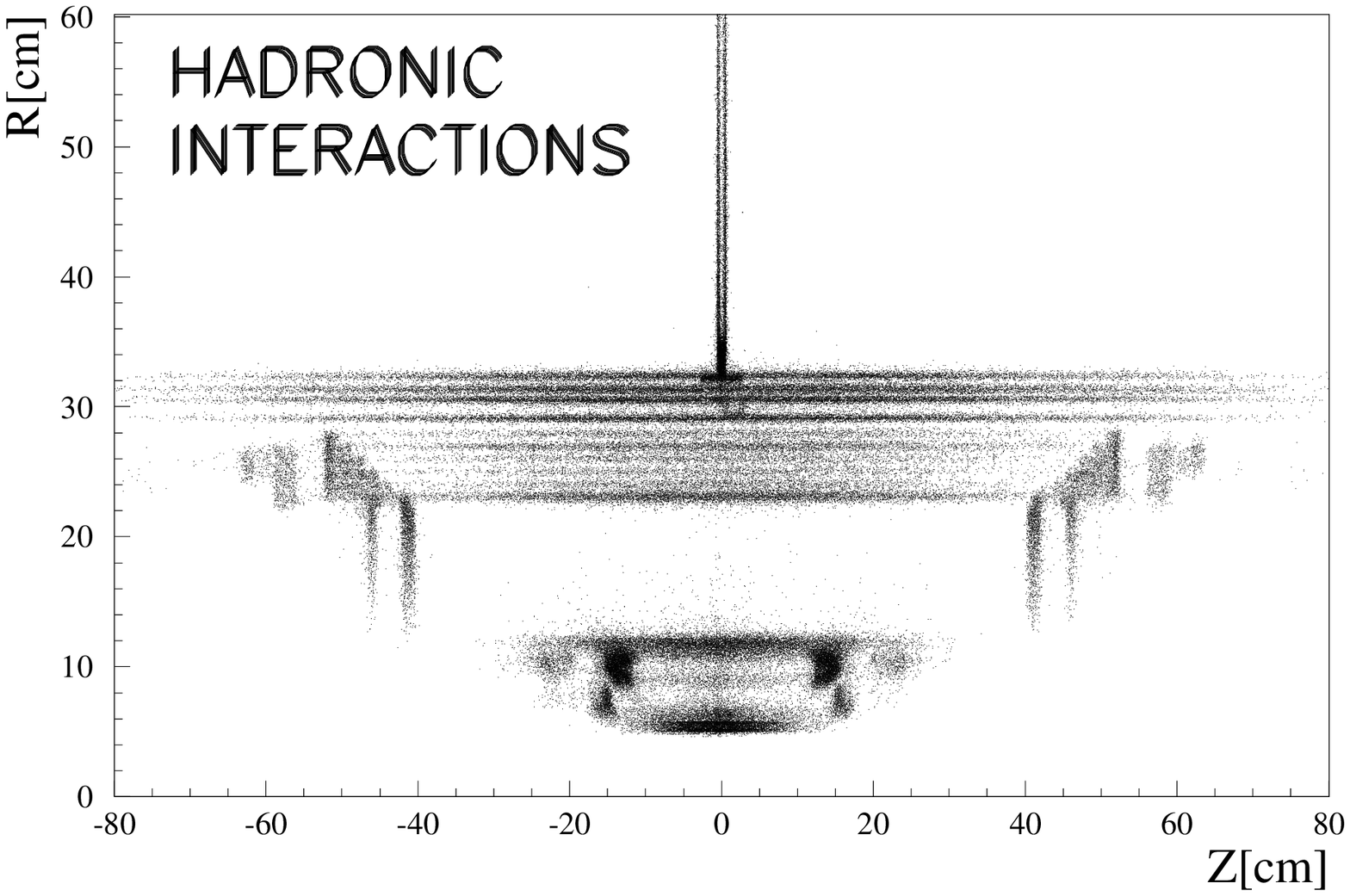} 
\end{flushleft}
\end{minipage}
\hfill  
\begin{minipage}[t]{0.39\textwidth}  
\begin{flushright}
\epsfig{bb=15 0 520 515,width=\textwidth,file=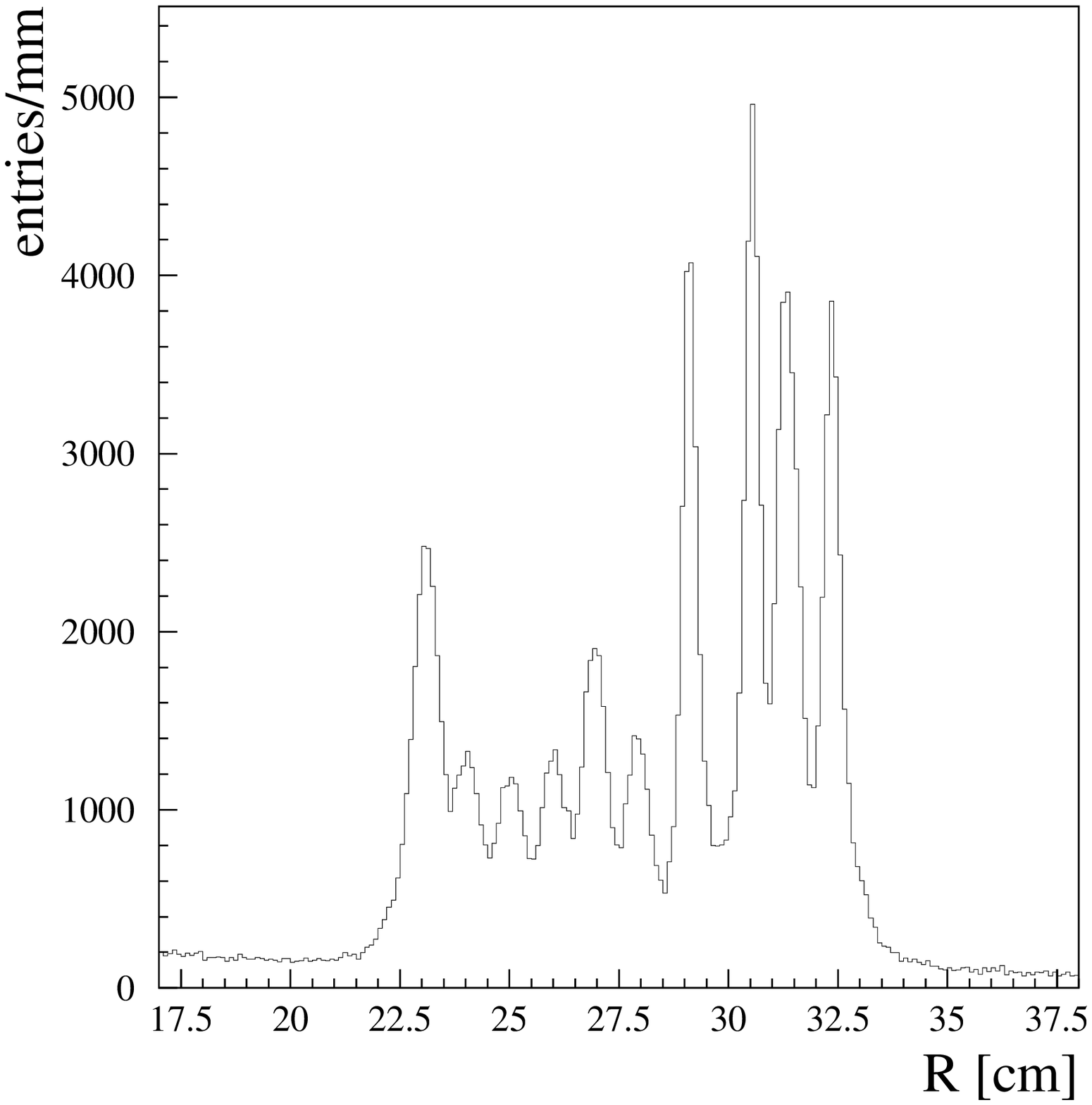}
\end{flushright}
\end{minipage}
\caption[]{\label{hadint}
         Reconstructed hadronic interactions in the material of 
         the \delphi detector, used
         for the determination of the correction factor for the VD track
         efficiency.
         Left: $Rz$ view; Right: Radial projection.}
\end{figure}

Hadronic interactions in the detector material with at least two outgoing
tracks were reconstructed by fitting the candidate tracks to a common vertex.
The algorithm allows for an arbitrary number and charge configuration of tracks
associated to the vertex\footnote{The charge of the outgoing tracks 
  may not sum up to the charge of the incoming track for hadronic interactions
  because an atomic nucleus is involved in the reaction.}.
The distribution of the measured positions of these vertices 
shows the material distribution of the \delphi detector (Figure \ref{hadint}).
If the incoming particle which caused the interaction is
charged\footnote{This is most often the case since the production rate of
neutral particles causing hadronic interactions (mainly K$^0_L$, n and
$\Lambda^0$) is small compared to the production rate of charged pions, kaons
and protons.}, there
is the possibility to reconstruct it as a VD track. 
To find these particles, 
the VD tracks were 
extrapolated to the radius of the interaction vertex and linked as the incoming
track to the vertex if the difference in the azimuth angle $\phi$
between the reconstructed vertex and the VD track  
was below 1.7\degr. This cut was chosen to achieve an efficiency and purity
for the linking close to unity. 
These linked VD tracks have not been used as
candidates for $\s$ hyperons to suppress background from hadronic interactions.
Finally, hadronic vertices within the polar angle
acceptance of the VD were selected and the number of interactions with a link
to a VD track was compared to the number of all selected interactions.
The efficiency for reconstructing a VD track is then given through
\begin{equation}
  \label{vdeff}
  \epsilon_{\mbox{\tiny VD}} = \frac{p_{hv/vd}\cdot N_{hv/vd}}
                                    {p_{hv}\cdot N_{hv}\cdot\epsilon_{link}\cdot f_c}
\end{equation}
where $N_{hv}$ is the number of hadronic vertices, $N_{hv/vd}$ the number of
hadronic vertices with a linked VD track, $p_{hv}$ and $p_{hv/vd}$ the purities
of these samples\footnote{$p_{hv}$ and $p_{hv/vd}$ were estimated directly
from the radial distribution of the vertices shown in Figure \ref{hadint}.},
$\epsilon_{link}$ the efficiency to link the incoming VD track
to the vertex and $f_c$ the fraction of charged particles causing hadronic
interactions in the detector material.
Assuming that the value of $f_c$ is
the same in real data and simulation and taking into account the
$\theta$ dependence by multiplying with the $\theta$ distribution
of $\s$ hyperons in simulated events, 
the following correction factor
for the reconstruction efficiency was deduced:
\begin{equation}
  \label{effratio}
  r = \frac{\epsilon_{\mbox{\tiny VD}}^{\mbox{\tiny RD}}}{\epsilon_{\mbox{\tiny VD}}^{\mbox{\tiny MC}}}
    = 0.91 \pm 0.06,   
\end{equation}
where the error is systematic and comes mainly from the uncertainty in the
fraction of charged particles causing hadronic interactions, which has been
estimated in the following manner.
In the region considered for the positions of reconstructed vertices
the fraction of charged particles causing the interactions in the
simulation was $f_c=0.84$.
As a conservative choice, the error on this fraction was chosen to cover the
range up to unity within three standard deviations, thus $f_c=0.84\pm 0.053$.
It is important to stress
that this method does not rely on a precise modelling of the material
distribution of the detector because the hadronic vertices were only
used as candidate endpoints for the VD tracks.

\subsubsection{Reconstruction of the decay {\boldmath{$\spmdec$}}}

The candidates for the outgoing pion had to fulfil the following criteria:
\begin{itemize}
\item Measured track length $>30$~cm.
\item $\Delta p/p<1$.
\item $IP/\sigma_{IP}>3$ in $R\phi$ and $z$,
      where $IP$ denotes the impact parameter with respect to the
      primary vertex and $\sigma_{IP}$ its error.
\item No associated VD hits in the two innermost layers.
\item The track must not originate from a reconstructed hadronic interaction.   
\end{itemize}
The $z$ information of the VD tracks with at least two VD hits was not used
in this analysis, because of differences in the association of two $z$ hits
between real data and simulation. Thus all VD tracks were
treated in exactly the same way.

\begin{figure}[tbh]
\begin{center}
\leavevmode 
\epsfig{bb=15 0 525 605,width=0.5\textwidth,file=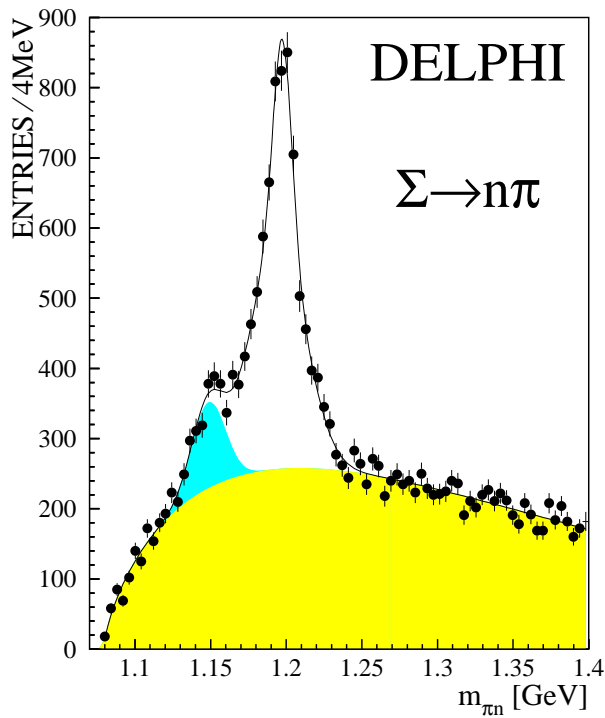}
\caption[]{\label{pinspectrum}
         The invariant mass spectrum for the $\s$ candidates selected as
         described in the text. Dots are the data.
         The curve shows the result of a fit with
         two Gaussians for the signal, one Gaussian for the reflection from the
         decay $\Xi^- \to \Lambda \pi^-$ (dark shading) and a function of the
         form (\ref{bgfunc}) for the smoothly varying background
         (light shading).
         The $\chi^2$ per degree
         of freedom of the fit is $\frac{83}{67}=1.24$.}
\end{center}
\end{figure}

To find the decay vertex of the $\spm$, the intersections (normally two) in the
$R\phi$ plane of the VD track with the previously selected pion candidates
were reconstructed. To select the correct intersection point, consistency of this
point with the incoming and outgoing track was required which included removing
intersections in the hemisphere opposite to the tracks. If both points fulfilled
all these cuts, the one with the lower radius was chosen. The $z$ coordinate of this
candidate decay vertex was given by the $z$ coordinate of the outgoing pion
candidate at this $R\phi$ position. The polar angle $\theta$ of the incoming track
was calculated using the $z$ coordinates of this vertex and of the primary
vertex completing the track parameters of the VD track.
The momentum of the (unreconstructed) neutron candidate could be computed
allowing the calculation of the invariant mass of the $\spm$-candidate.
The combination was rejected if the calculated $\theta$ did not
lie within the polar angle range covered by the modules of the VD
defining the VD track.
Both tracks defining the vertex must have the same charge. In order to
reject background, 
the following additional cuts were applied:
\begin{itemize}
\item The probability of the particle decaying within the measured flight distance,
      calculated under the $\sm$ hypothesis, had to be lower than 97\%.
      This efficiently removes background from the decays
      $\mbox{K}/\pi \to \mu \nu_{\mu}$. Due to their long
      lifetime ($c\tau$(K$^{\pm}) \approx$ 3.7 m
      and $c\tau(\pi^{\pm}) \approx$  7.8 m) charged kaons and pions 
      decaying inside the sensitive volume of the tracking chambers tend to have
      very low momenta. Thus, their decay probability as defined above is
      close to 100\%.
\item $|\cos\theta^*| < 0.8$, 
      where $\theta^*$ denotes the angle between the outgoing track
      and the VD track, calculated in the rest frame of the particle
      reconstructed as
      VD track, where the pion mass was assumed for the
      measured outgoing track and the neutron mass for the
      undetected particle.
\end{itemize}
The resulting mass spectrum of the $\s$ candidates is shown in
Figure \ref{pinspectrum}. 
A fit with Gaussians for the signal and the reflection from
$\Xi^- \to \Lambda\pi^-$ ($c\tau = 4.91$ cm), which has the same
signature of a kink\footnote{The decays $\Omega^-\to\Lambda\mbox{K}^-,\Xi^0\pi^-$
  also show this signature, but their contribution is negligible due the low
  $\Omega^-$ production rate.},
and a function of the form
\begin{equation}
  \label{bgfunc}
  F(M) = a_1 ( (M-1.079) (a_4-M) )^{a_2} \exp(-a_3(M-0.5(1.079+a_4)))
\end{equation}
for the background, where 1.079 is the sum of the masses of the neutron
and the pion (in GeV/$c^2$) and thus the lower kinematical limit,
gives a mean mass of $1196.5 \pm 0.4$~MeV/$c^2$ and a width 
$\sigma=12.3 \pm 0.4$~MeV/$c^2$, in good agreement with the expectation from simulation
of $1197.4 \pm 0.3$~MeV/$c^2$ and $12.2 \pm 0.6$~MeV/$c^2$, respectively.
To subtract the background, two Gaussians with all parameters left free
have been used for the signal to
take into account the momentum dependence of its width.
This results in a measured signal of 
$4820 \pm 109$ $\s$ decays and $870 \pm 95$ $\Xi^-$ decays
(statistical errors only).
\subsubsection{Measurement of the {\boldmath{$\sm$}} differential production rate}

Due to the larger branching ratio into the final state n$\pi$
(see Eq. (\ref{sigmaprop})) and the longer lifetime
($c\tau_{\sm}\approx 1.8\cdot c\tau_{\sp}$),  
the efficiency to reconstruct the decay $\sdec$ with
the method described above is much higher for the $\sm$ than for the $\sp$.
To obtain the $\sm$ cross-section
\begin{displaymath}
  \sigma_{\sm} \sim \frac{N}{\epsilon^-} - \frac{1}{a}N_{\sp}
\end{displaymath}
where $N$ denotes the number of signal events, $\epsilon^-$ the
efficiency for $\sm$ reconstruction, $a=\frac{\epsilon^-}{\epsilon^+}$ the ratio
of the efficiencies for detecting a $\sm$ or $\sp$,
and $N_{\sp}$ the true number of $\sp$ in the signal, an estimate for the
$\sp$ rate to be subtracted has to be made.
For this, the ratio of $\sp$ to $\sm$ production from the simulation
(\jsf with default parameters) has been assumed. An error of 20\% has been assigned
to this ratio. In the $x_p$ bins ($x_p=2p/\sqrt{s}$)
where the deviation of this ratio from unity
is greater than 20\%, this deviation has been taken as the systematic error.

\begin{figure}[th]
\begin{center}
\leavevmode 
\epsfig{bb=30 30 380 460,width=0.7\textwidth,file=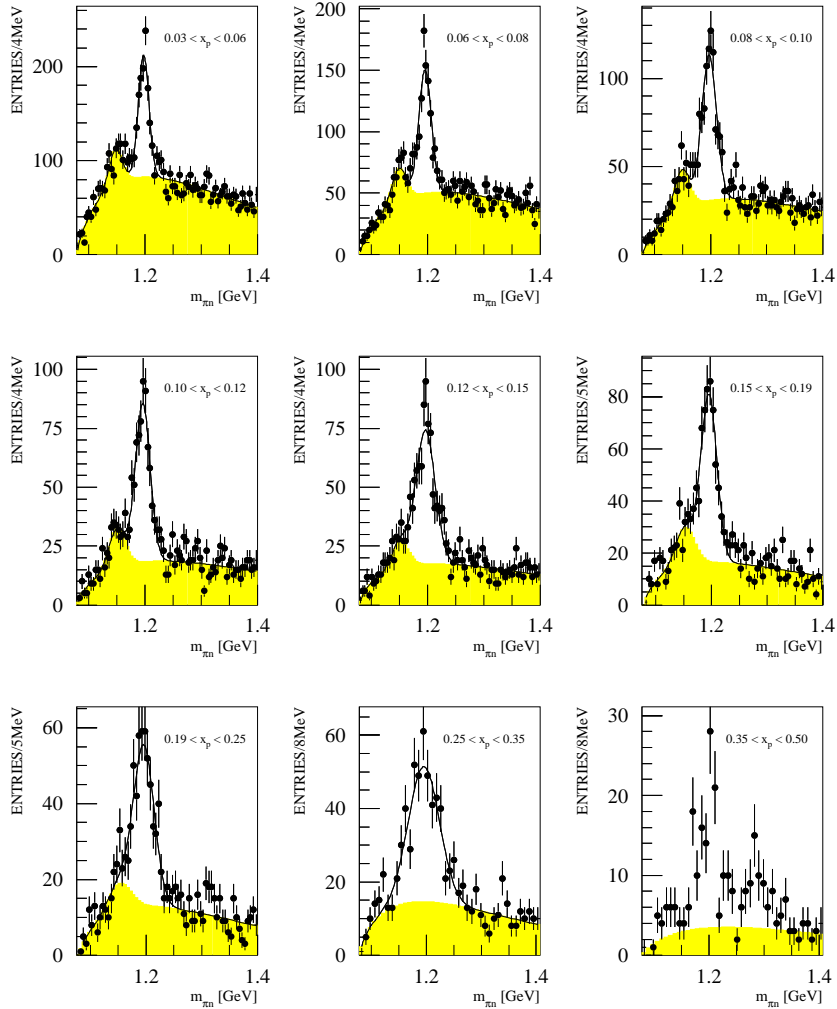}
\caption[]{\label{allxpspectra}
         The mass spectra for the different $x_p$ bins.
         Dots are the data, the solid line
         shows the result from the fit and the shaded histogram the
         background used for subtraction.}
\end{center}
\end{figure}
The differential $\sm$ production rate has then been measured in
nine $x_p$ bins.
For each $x_p$ bin the efficiency has been estimated using the
simulated sample with full detector simulation,
taking into account the correction factor of
equation (\ref{effratio}) which was assumed to be independent of $x_p$.
To obtain the number of signal events, the
mass spectra have been fitted using a Gaussian function for the signal,
a Gaussian function with position and width fixed to the values obtained
from the simulation for the reflection $\Xi^- \to \Lambda \pi^-$,
and a function with four free parameters
of the form (\ref{bgfunc}) for the smoothly varying
background (see Figure \ref{allxpspectra}).
The reflection has not been fitted separately for the last two
$x_p$ bins due to the large width of the signal. In the last $x_p$ bin the
signal has not been fitted, but the
background obtained from the simulation
(including the contribution from the $\Xi$-reflection and
normalised to data statistics)
has been subtracted to get the number of signal events.

\subsection{Consistency checks and systematic errors}

In addition to the systematic error already mentioned in Section \ref{s:effcorr},
some more systematic checks have been done to
test the consistency between real data and simulation.
The quantities used for the selection of the candidates have been
compared and good agreement was found.
The mass spectrum has been fitted separately for negative 
and positively charged $\s$ candidates. One obtains
good agreement within the statistical errors
($2351 \pm 73$ and $2298 \pm 78$ signal events, respectively).
Since $a=\frac{\epsilon^-}{\epsilon^+}$ (the values are given
in Table \ref{tableresults}) is a function of the decay length of
$\sm$ and $\sp$, the flight distance distributions for real data and
the simulation have been compared for each $x_p$ bin and consistency
within the statistical errors has been found.

Since this analysis uses tracks which have been reconstructed
using only the microvertex detector of \delphi, a good internal and external
alignment of the VD is essential. 
It has been checked that the widths obtained for the signal in the
different $x_p$ bins
show good agreement between real data and simulation
within errors. The signal has been fitted separately for
both $z$ hemispheres defined through $\cos \theta_{VD-track} \gtrless 0$.
Taking into account the $\theta$ dependence of the efficiency for
reconstructing a VD track (according to (\ref{vdeff})), one obtains
$N_{\cos\theta<0} / N_{\cos\theta>0} = 1.05 \pm 0.05 \mbox{ (stat.)}$.
Another quantity sensitive to the alignment and effects of the tracking,
such as misassociating VD hits to the outgoing pion, is the kink angle
between the two tracks. The simulation describes the
data well, even for very low values ($\lesssim$ 5\degr) of this angle,
where the reconstruction
of two distinct tracks and the kink between them is most critical.

To estimate the uncertainty of the number of signal events, the
parametrisation of the background and the mass window for the
background subtraction have been varied.

The systematic errors due to the efficiency correction and the subtraction of
the $\sp$ rate have been added linearly for the different $x_p$ bins
and thus treated as fully correlated.
Statistical and other systematic errors have been added quadratically from
bin to bin.
Different systematic errors have been added
quadratically. The contributions from the different sources to the total error
are listed in Table \ref{errors}.
%
%
\begin{table}[h]
\begin{center}
\begin{tabular}{|l|c|c|}\hline
error source                      & absolute unc. & relative unc. \\ \hline
data statistics                          & 0.0016 & 2.0\% \\ \hline
simulation statistics                    & 0.0012 & 1.5\% \\
efficiency correction
    (equation (\ref{effratio}))          & 0.0043 & 5.3\% \\
fit procedure                            & 0.0020 & 2.5\% \\
$\sp$ rate from simulation               & 0.0028 & 3.5\% \\
extrapolation to unobserved $x_p$ region & 0.0081 & 10.0\% \\ \hline
\end{tabular}
\end{center}
\caption[]{\label{errors}
         The uncertainties of the $\sm$ production rate.}
\end{table} 


\subsection{{\boldmath{$\lft$}} Reconstruction \label{s:l1520rec}}
Beyond the general cuts given in Section \ref{s:det_ana} for this analysis
it was required that the track impact parameter to the primary vertex
was less than 0.5~mm in the $R\phi$-plane and 1~mm in the $z$-direction. 
This requirement strongly reduces contributions of tracks from particle
re-interactions inside the detector material.
Furthermore there must
be at least two tracks inside the angular acceptance
$47^{\circ}<\theta<133^{\circ}$ of the BRICH.
\newcommand{\mc}{\multicolumn}
\begin{table}[tbh]
{
\begin{center}{
\begin{tabular}{|c|c|c|c|c|c|c|c|}
   \hline
   & \multicolumn{7}{c|}{momentum range in~$\gev/c$}\\
  \cline{2-8}
   & 0.3 - 0.7 & 0.7 - 0.9 & 0.9 - 1.3 & 1.3 - 2.7 & 2.7 - 9.0 &
                                         9.0 -16.0 & 16.0 - 45.0 \\
  \hline
  \hline
       &     & \mc{3}{c|}{}        & \mc{3}{c|}{}\\
 $\pi$ & TPC & \mc{3}{c|}{LRICH S} & \mc{3}{c|}{GRICH S}\\
       &     & \mc{3}{c|}{}        & \mc{3}{c|}{}\\
  \hline
       &     & \mc{3}{c|}{}        & GRICH V &\mc{2}{c|}{}\\
   K   & TPC & \mc{3}{c|}{LRICH S} &    +    &\mc{2}{c|}{GRICH S}\\
       &     & \mc{3}{c|}{}        & LRICH S &\mc{2}{c|}{}\\
  \hline
       & \mc{2}{c|}{}    &  TPC    &         & GRICH V &         &         \\
   p   & \mc{2}{c|}{TPC} &  +      & LRICH S &    +    & GRICH V & GRICH S \\
       & \mc{2}{c|}{}    & LRICH V &         & LRICH S &         &         \\
  \hline
\end{tabular}}
\end{center}
\caption[Impulsabh"angige Einsatzbereiche der Detektoren
                                       zur Teilchenidentifikation]
        {Momentum ranges for particle identification: TPC denotes
         identification using the $dE/dx$ measurement of the TPC, 
         LRICH S (V) denotes identification using a signal (veto) of
         the liquid RICH, and correspondingly GRICH for the gas RICH.
}
\label{t:id-ranges}
}
\end{table}

The $\rm pK^{-}(\bar{p}K^+)$ mass-spectra were then constructed 
for each bin of each individual kinematical variable using identified particles.
Particle identification was performed combining $dE/dx$ and BRICH information.
According to the quality of particle identification 
the tagging categories
loose, standard and tight tags are 
distinguished for each particle species as well as for so called
``heavy'' particles combining protons and kaons.
To further improve the quality of particle identification for a track of given
momentum and (assumed) particle type it was required that information 
from the detectors specified in Table~(\ref{t:id-ranges}) was present.

A particle was then taken to be a proton if it was tightly tagged.
Kaons were required to be tightly tagged in the momentum ranges 
$p<3.5$~\gev$/c$ and $p>9.5$~\gev$/c$. 
In the
intermediate momentum range kaons were also identified by a tight heavy
particle tag combined with at least a standard kaon tag.
To suppress combinatorial background 
it was required that the kaon momentum was between 28\% and 100\% of the proton
momentum.
This condition has been left out for the determination of the $\lft$ spin alignment.
\newlength{\wi}   \wi 8cm
\newlength{\fwi}   \fwi 8cm
\begin{figure}[tbh]
\fwi 0.60\textwidth
\center{\epsfig{file=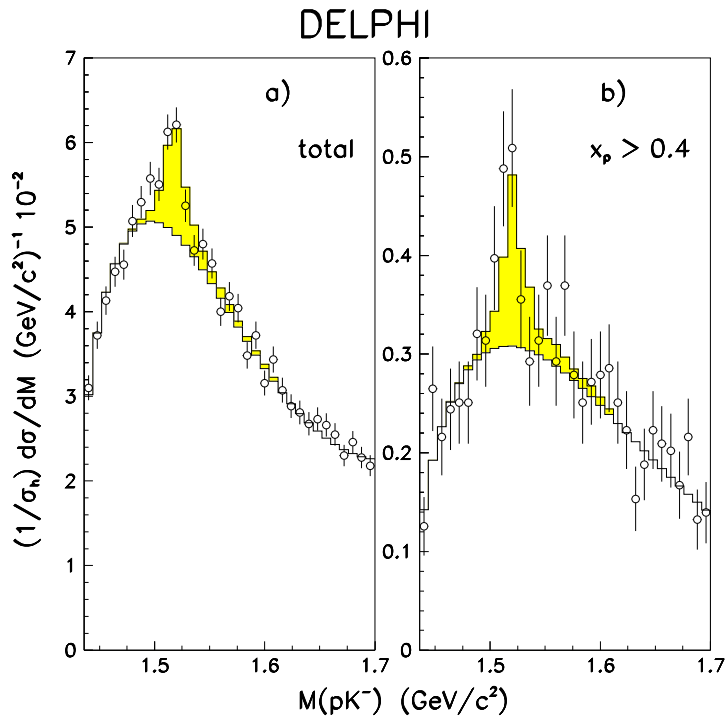,width=\fwi}}
\caption[massplots l1520]
{
\label{f:l1520mass}
a) Differential pK$^-$ mass spectra for the overall measured energy range.
b) pK$^-$ mass spectrum for $x_p>0.4$.
The histograms represent the fit described in the text.
}
\end{figure}

Figure (\ref{f:l1520mass}a) shows the pK$^-$ mass spectrum of 
the overall dataset and Figure (\ref{f:l1520mass}b)
for the scaled momentum range $x_p>0.4$.
A clear $\lft$ signal is observed in both mass spectra at about the expected mass.
The signal to noise ratio improves for the higher $x_p$ range indicating
that a proper measurement can even be performed for high $\lft$ momenta.
It should, however, be noted that here the $\lft$ signal became poorly visible 
if the particle identification requirements were relaxed.
It has been checked that there are no prominent reflections from known
particle decays in the pK$^-$ mass spectrum.

Particle identification inefficiencies, detector imperfections 
and the different kinematical 
cuts imposed for charged particle and event selection, were accounted 
for by applying the approach first described in \cite{d_ro}, developed 
in \cite{d_kot,d_del,spin,d_zus} and outlined in brief below. 

In the present analysis, the mass spectra were described by an anticipated 
distribution function,  $f(M,\vec{a})$, of the invariant mass $M$.
The parameters $\vec{a}$
were determined by a least squares fit of the function to the data.
The function $f(M,\vec{a})$ had three components:
\begin{equation}
   f(M,\vec{a}) = f^{S}(M,\vec{a}) + f^{B}(M,\vec{a}) 
                                     + f^{R}(M,\vec{a}),
\end{equation}
corresponding to the signal ($S$), background ($B$), 
and reflection ($R$) contributions respectively.

The signal function, $f^{S}(M,\vec{a})$, described the resonance signal 
in the corresponding invariant mass distributions. For the pK$^-$ mass 
distributions it had the form
\begin{eqnarray}
f^S(M,\vec{a}) = a_1 PS(M)\cdot BW(M,a_{2},a_{3}),
\end{eqnarray}
where the relativistic Breit--Wigner function $BW$ for the $\Lambda$(1520)
is multiplied by the function $PS(M)$ to
account for the distortion of the resonance Breit--Wigner shape 
by phase space effects (see \cite{d_ro} for details). 

The background term, $f^{B}(M,\vec{a})$, was taken to be of the form
\begin{equation}
 f^{B}(M,\vec{a}) = BG_{Jetset}(M)\cdot P(M,\vec{a}),
\end{equation}
where $BG_{Jetset}(M)$ represented the background shape generated 
by JETSET which describes the gross features of the real 
background \cite{fred} and $P(M,\vec{a}) = 1 + a_{4}M +a_{5}M^2 + a_{6}M^3 + 
a_{7}M^4$ was a polynomial of order four (sometimes $a_7$ has been fixed to zero) 
introduced to account for possible deviations of $BG_{Jetset}(M)$ 
from the real background. 
All pairs of charged particles which do not come from the resonance
considered and reflections in the invariant mass spectra were 
included in the definition of $BG_{Jetset}(M)$.
This parametrisation of the background was different from the 
analytical form used in a previous 
DELPHI analysis~\cite{d_ro,d_kot,d_del,spin}.

The third term, $f^{R}(M,\vec{a})$, represented 
the sum of all the reflection functions ($RF_i$):
\begin{equation}
 f^{R}(M,\vec{a}) = {\sum_i} {a_{i+7}RF_{i}(M)}.
\label{f:fM}
\end{equation}
Reflections arising from particle misidentification and contributing to 
Equation~(\ref{f:fM}) 
were considered, for example when resonances in the $\pi^+\mbox{K}^-$,
$\mbox{K}^+\mbox{K}^-$ and 
p$\pi^-$ systems (K$^{*0}, \Phi, \Delta^0$) distort the pK$^-$ mass spectra. 
Due to the efficient particle identification of the combined 
RICH and TPC tags and to the high identification purity provided by the 
tight cuts, the influence of reflections of this
type was found to be much smaller than without particle identification. 

The functions $RF_i(M)$ in Equation~(\ref{f:fM}) were determined from events 
generated according to the JETSET model. The contributions of
the reflections to the raw mass spectra defined by the 
function ${\bar N}^R_m(\vec{a})$ (see Equation~(\ref{f:Nm}) ) were then 
obtained by passing these events through the detector simulation. 
This also took proper account of the influence of particle 
misidentification.

In each mass bin, $M$, the number of entries ${\bar N}_m(\vec{a})$ 
predicted by the function $f(M,\vec{a})$, representing a sum of 
contributions from the resonance signal, background and reflections 
(see \cite{d_zus}), is given by 
\begin{equation}
{\bar N}^G_{m}(\vec{a}) = 
    C^G_m \sum_n{S^G_{mn} A^G_n f^G_{n}(\vec{a})},
\label{f:Nm}
\end{equation}
\begin{equation}
f^G_{n}(\vec{a}) = \int_{M_{n}}^{M_{n+1}} f^G(M,\vec{a}) dM,
\end{equation}
where $G = S$, $B$ or $R$, and $M_{n}$ is the lower edge of the $n$-th 
histogram bin in the distribution of the variable $M$. The coefficients $A_n$ 
characterise the detector acceptance and the losses of particles 
due to the selection criteria imposed, and the $C_m$ take into account 
the contamination of the sample by particles from $V^0$ decays, 
wrongly associated charged particles, secondary interactions, etc.
The smearing matrix $S_{mn}$ represents the experimental 
resolution. The $A_n$, $C_m$ and $S_{mn}$ were estimated separately 
for the resonance signal, background and reflection contributions
using the detector simulation program.
Due to differences in 
the detector performance and data processing in different running 
periods, the simulated events generated for these periods were taken
with weights corresponding to the relative number of
events in the real data. 

The best values for $\vec{a}$ were then determined by a least squares
fit of the predictions of Equation~(\ref{f:Nm}) to the measured values, $N_m$, 
by minimising the function
\begin{equation}
   \chi^{2} = \sum_{m}{(N_{m}-\bar N_{m}(\vec{a}))^{2}/\sigma^2_m}
   +{\sum_{i}}{(a_{i}-\bar a_{i})^2/(\Delta\bar a_{i})^2},
\label{f:chi2}
\end{equation}
where $\sigma^2_m = N_m + \sigma^{2}(\bar {N}_{m})$ and
$\sigma(\bar {N}_{m})$ is the error on $\bar {N}_m$ due to the 
finite statistics of the simulation used to evaluate $A_n$, $C_m$ 
and $S_{mn}$. The second sum in Equation~(\ref{f:chi2}) constrains some of the fitted 
parameters $a_i$ to the values $\bar a_{i} \pm \Delta\bar a_{i}$ 
taken from external sources, such as the normalisation of the 
reflection functions to the particle production rates taken from 
this and other LEP experiments, and the mass and width taken 
from the PDG tables~\cite{PDG}. The errors obtained from the 
fits thus include the corresponding systematic components.
As a cross-check the $\lft$ mass was also left free in the fit.
This lead to a mass of $1517.5 \pm 1.7$ \mev$/c^2$, fully consistent
with the PDG value. 

In order to determine the full experimental systematic error of the $\lft$
cross-section the following sources of uncertainty were considered.
The influence of an imperfect description of stable hadron spectra by 
the detector simulation was estimated
by varying the charged particle selections leading to an associated systematic 
error of 3\%.
An error of 9\% was assigned due to the imperfect description of $\lft$ 
production by the fragmentation model.
It was estimated by varying kinematical distributions of the $\lft$ like the 
decay angle and momentum distribution.
An uncertainty of 3\% on the resonance cross-sections is due to the 
imperfect description of the particle identification 
efficiency\cite{EPJC5_585} and the
error due to the branching ratios assumed is 2.2\% \cite{PDG}.
Uncertainties due to the unknown momentum dependence of the reflections 
(3\%) were
assessed by reweighting the shape of the momentum distribution predicted by 
the model in a range allowed by the fit.
Finally the uncertainty in the $\lft$ line-shape, the background parametrisations 
and the influence of the binning was estimated to be 12\%
by choosing different parametrisations 
and by changing the binnings of the mass spectra.
Adding the individual contributions in quadrature leads to a final relative
experimental error of the $\lft$ cross-section of 16\%.

\section{Results 
\label{s:results}}

\subsection{\boldmath{$\sm$}}

\begin{figure}[t]
\begin{center}
\leavevmode 
\epsfig{bb=35 25 515 655,width=0.6\textwidth,file=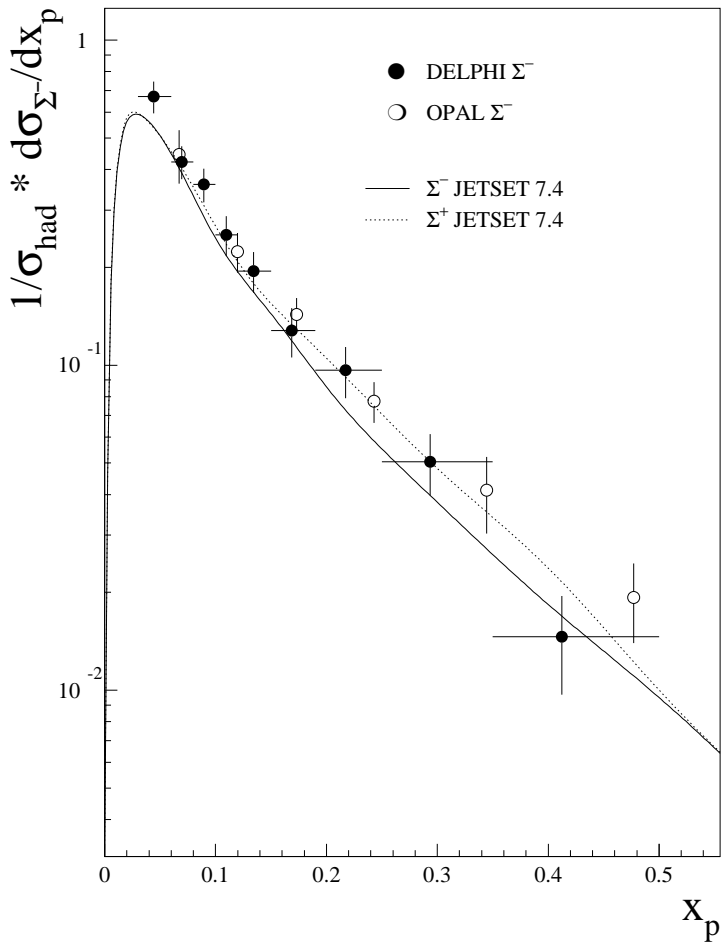}
\caption[]{\label{sigmacross}
         The measured differential $\sm$ production rate in comparison
         with the \jetset model and
         the \textsc{OPAL} measurement \cite{opalsigma}.
         The data points are plotted at the mean $x_p$ position in
         the corresponding bin.
         The statistical and systematic errors have been added
         quadratically.}
\end{center}
\end{figure}

The results for the differential $\sm$ production rate and reconstruction efficiencies
are listed in Table \ref{tableresults}.
The differential $\sm$ production rate is plotted in Figure \ref{sigmacross}
together with the predictions of the \jsf (default parameters) generator and
a previous measurement of the \textsc{OPAL} collaboration \cite{opalsigma}.
The higher $\sp$ rate compared to the $\sm$ rate
in \jetset is due to secondary $\s$ hyperons, mainly from decays of
charm particles.
The shape of the $x_p$ spectrum is well described in the simulation.
%
%
\begin{table}[bht]
\begin{center}
\begin{tabular}{|c||c|c|c|c|c|}\hline
   & & & & & \\
 $x_p$ & $\epsilon^-$(\%) & $a=\frac{\epsilon^-}{\epsilon^+}$ & $\frac{\chi^2}{\mbox{n.d.f.}}$ & $N_{Signal}$ & $\frac{1}{\sigma_{had}}\frac{d\sigma_{\sm}}{dx_p}$ \\ 
   & & & & & \\ \hline\hline
 0.03 - 0.06 & 2.8 $\pm$ 0.1 & 6.5 & 1.7 & 860 $\pm$ 52 & 0.671 $\pm$ 0.041  $\pm$ 0.064 \\     
 0.06 - 0.08 & 5.6 $\pm$ 0.2 & 4.5 & 1.2 & 745 $\pm$ 44 & 0.422 $\pm$ 0.025  $\pm$ 0.042 \\
 0.08 - 0.10 & 5.9 $\pm$ 0.2 & 3.4 & 1.1 & 725 $\pm$ 40 & 0.361 $\pm$ 0.020  $\pm$ 0.038 \\ 
 0.10 - 0.12 & 6.5 $\pm$ 0.3 & 3.4 & 1.3 & 560 $\pm$ 34 & 0.252 $\pm$ 0.015  $\pm$ 0.032 \\
 0.12 - 0.15 & 6.2 $\pm$ 0.2 & 3.2 & 1.1 & 630 $\pm$ 35 & 0.195 $\pm$ 0.011  $\pm$ 0.026 \\
 0.15 - 0.19 & 5.0 $\pm$ 0.2 & 2.7 & 1.4 & 470 $\pm$ 32 & 0.128 $\pm$ 0.009  $\pm$ 0.020 \\
 0.19 - 0.25 & 4.1 $\pm$ 0.2 & 2.7 & 1.2 & 450 $\pm$ 29 & 0.097 $\pm$ 0.006  $\pm$ 0.016 \\
 0.25 - 0.35 & 3.7 $\pm$ 0.2 & 2.2 & 1.2 & 380 $\pm$ 26 & 0.051 $\pm$ 0.004  $\pm$ 0.010 \\
 0.35 - 0.50 & 4.0 $\pm$ 0.3 & 2.2 & --- & 175 $\pm$ 17 & 0.015 $\pm$ 0.001  $\pm$ 0.005 \\ \hline
\end{tabular}
\end{center}
\caption{\label{tableresults} 
         Efficiencies and differential $\sm$ production rate in bins of $x_p$.
         The errors given for $\epsilon^-$ and $N_{Signal}$ are coming
         from the simulation
         and data statistics, respectively.}
\end{table} 

Integrating the differential production rate in the measured $x_p$
range from 0.03 to 0.5 gives
\begin{displaymath}
  \langle N_{\sm}/N_{\rm Z}^{\rm had}\rangle = 0.065 \pm 0.002\mbox{ (stat.)}
                                           \pm 0.006\mbox{ (syst.)} \quad \mbox{(\jsf: 0.054)}.
\end{displaymath}

To get the total production rate, the \jsf simulated data has been
used to extrapolate to the unobserved $x_p$ range.
In \jsf, 18\% of the $\sm$ hyperons
are produced with $x_p < 0.03$ and 2\% with $x_p > 0.5$.
The production cross-section in this $x_p$ range has been scaled
by the ratio of the measured and simulated cross-section in the measured
range $0.03<x_p<0.5$. This scaled cross-section has then been added and a
systematic error of 50\% has been assigned to it.

This gives the mean number of $\sm$ hyperons produced in multihadronic
Z$^0$ decays
\begin{displaymath}
  \langle N_{\sm}/N_{\rm Z}^{\rm had}\rangle = 0.081 \pm 0.002\mbox{ (stat.)}
                                           \pm 0.006\mbox{ (syst.)}
                                           \pm 0.008\mbox{ (extr.)},
\end{displaymath}
where the last error comes from the extrapolation.

This result is compatible
with the corresponding rate
in \jsf ($\langle N_{\sm}/N_{\rm Z}^{\rm had}\rangle = 0.068$).
It is also in good agreement with
the measurement from \textsc{Opal} \cite{opalsigma}:
$\langle N_{\sm}/N_{\rm Z}^{\rm had}\rangle = 0.083 \pm 0.006\mbox{ (stat.)}
                                          \pm 0.009\mbox{ (syst.)}$.


\subsection{\boldmath{$\lft$}}
The $\lft$ has $J^P = \frac{3}{2}^-$ with isospin 0 and quark content (uds).
It decays strongly but with a comparably small decay width of $15.6\pm1.0$~MeV/$c^2$
as it is a $D_{03}$ state which predominantly decays into 
a $J^P = \frac{1}{2}^+$ baryon and one pseudoscalar meson.
The $\lft$ branching fraction used in the analysis is taken as half of the 
branching fraction to nucleon and kaon of 45\%:
\begin{eqnarray*}
 B( \lft & \rightarrow \mbox{pK}^-) & \simeq 22.5\% ~~~.
\end{eqnarray*}

The total $\lft$ rate is measured from a fit to the mass spectrum given in
Figure~(\ref{f:l1520mass}a) corresponding to the
scaled energy\footnote{$x_E^{\lft}=\frac{2E^{\lft}}{\sqrt{s}}$}
range $0.035 < x_E^{\lft} < 1$:
\begin{equation*}
\langle N_{\lft} / N_{\rm Z}^{\rm had} \rangle = 0.0285 \pm 0.0048 \mbox{ (fit)}~~~.
\end{equation*}
The fit error includes the statistical error and also accounts for
uncertainties in the $\lft$ mass and width and the normalisation of the
reflection functions (see Section \ref{s:l1520rec}).
The fitted rate agrees well with the integrated rate from the $\lft$
inclusive $x_E$ spectrum (see below).
To estimate the total rate of $\lft$ production this value has to be corrected
for the small unmeasured $x_E$ range $x_E^{\lft}<0.035$.
This correction was estimated using a modified \jsf model 
(see below)
normalised to the data in the measured momentum range. 
An additional systematic
error of 50\% of this correction has been assumed. This yields for the overall
$\lft$ production rate
\begin{equation*}
\langle N_{\lft} / N_{\rm Z}^{\rm had} \rangle = 0.0293 \pm 0.0049 \mbox{ (fit)} 
           \pm 0.0047 \mbox{ (syst.)} \pm 0.0003 \mbox{ (extr.)}
\end{equation*}
including the systematic errors considered in Section \ref{s:l1520rec}.
This result is slightly higher than, though fully consistent with, the \opal result
\cite{opal_res_paper}. 

In order to exclude that the observed $\lft$'s originate predominantly 
from heavy (b,c) particle decays
the $\lft$ production has been determined individually for samples 
strongly enriched in b-quarks and light quarks respectively \cite{fred}. 
This enrichment relies on
increased particle impact parameters due to the high lifetime of 
B hadrons \cite{btag-ref}.
No significant change in $\lft$ production has been found in either sample 
which leads to the
conclusion that the dominant part of $\lft$ production originates 
from fragmentation.

The observed $\lft$ rate is comparable to that of the
$\Sigma^{*\pm}(1385)$ \cite{opal_res_paper,sigmaolddelphi,alephlambda} 
which has the same strangeness and total spin. 
It can be concluded
that the orbital excitation ($L=1$) of the $\lft$ does not 
lead to a suppression of particle production. 
The comparably large observed $\lft$ rate
suggests that also other orbitally excited baryonic states are produced in fragmentation.
In consequence, due to the vast amount of orbitally excited states, a large part
of the observed stable baryons may descend from these excited states.
This agrees with the expectations of the phenomenological 
models~\cite{chliapnikov,pei}.
Thus for baryons the situation is likely to be similar to the mesonic case.
It is highly desirable to verify the production of other orbitally
excited baryonic states, however, this is experimentally 
demanding due to the large width and complicated decay modes of these states.

\begin{figure}[tbh]
\wi 0.6\textwidth
\center{\epsfig{file=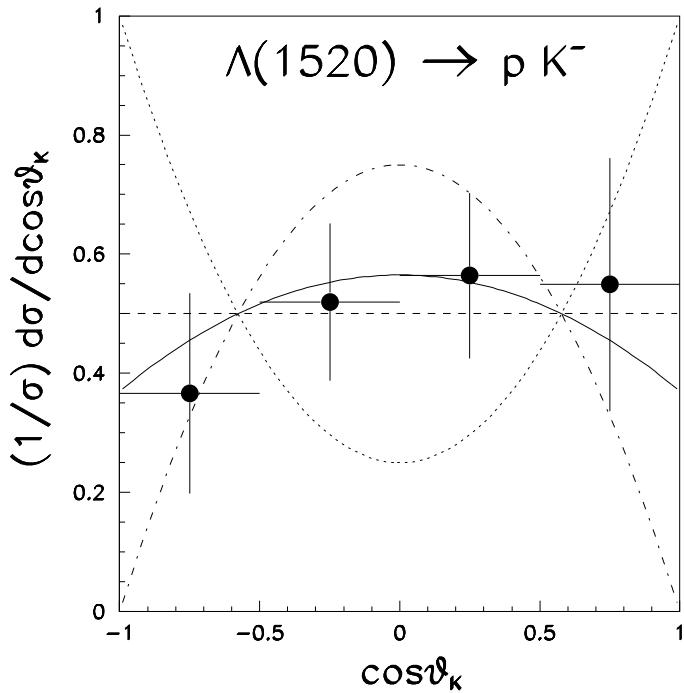,width=\wi}}
\caption[costh l1520]
{
\label{f:l1520angle}
Differential $\lft$ distribution as function of $\cos\vartheta_K$ 
for $x_p > 0.07$.
The full line represents the result of a fit of the angular distribution,
the dashed line represents the expectation for unaligned $\lft$'s and the dotted
(dash-dotted) line represents the expectation for 
$\rho_{\frac{1}{2}\frac{1}{2}} + \rho_{-\frac{1}{2}-\frac{1}{2}}=1~(0)$.
}
\end{figure}

In order to determine a possible spin alignment of the $\lft$'s the distribution
of the cosine of the kaon angle, $\cos\vartheta_K$,  in the $\lft$ rest system 
with respect to the $\lft$ direction
is plotted in Figure (\ref{f:l1520angle}) for $x_p > 0.07$. 
This distribution has been fitted with the expected form of the angular
distribution:
\begin{equation}
W(\cos{\vartheta_{\rm K}})=\rho\frac{1+3\cos^2{\vartheta_{\rm K}}}{4}+
(1-\rho)\frac{3-3\cos^2{\vartheta_{\rm K}}}{4}
\end{equation}
with $\rho = \rho_{\frac{1}{2}\frac{1}{2}} + \rho_{-\frac{1}{2}-\frac{1}{2}}$,
$\rho_{ii}$ denoting the spin-density matrix element.
The fit yields $\rho=0.4\pm0.2$, thus
no significant $\lft$ spin alignment is observed. 
The specified error includes the statistical error and the error due to the
uncertainty in the background shape.

Furthermore, the approximate symmetry 
of this distribution with respect to $\cos\vartheta_K=0$ 
presents evidence for the validity of this
analysis as possible errors in particle identification and corrections would,
in general, lead to a distorted $\cos\vartheta_K$ distribution.

\begin{figure}[tbh]
\wi 0.6\textwidth
\center{\epsfig{file=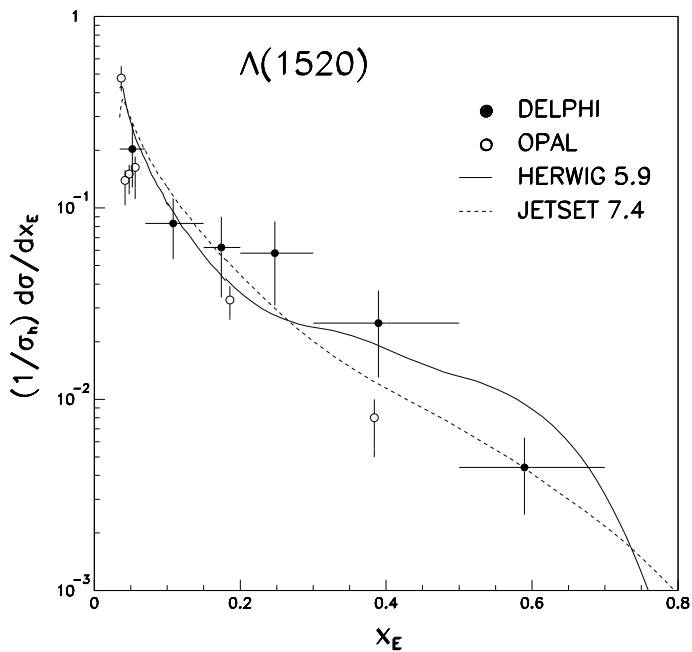,width=\wi}}
\caption[costh l1520]
{
Inclusive $\lft$ distribution as function of $x_E$.
Horizontal error bars indicate the bin width, vertical error bars are the fit errors
described in the text.
\label{f:l1520dndxe}
}
\end{figure}

\begin{table}[tbh]
\begin{center}
\begin{tabular}{|c|c|ccccc|c|}
\hline
  $x_E$-range  & $\langle x_E \rangle$ &  \mc{5}{c|}{
                                      $1/N_{evt.}\cdot dN^{\lft}/dx_E$}  &
                                                 $\chi^2/N_{df}$ \\ \hline
0.035~--~0.07 & 0.052 & 0.203  & $\pm$ & 0.068  & $\pm$ & 0.032  &  28 / 27  \\
0.07~~--~0.15 & 0.108 & 0.083  & $\pm$ & 0.026  & $\pm$ & 0.013  &  32 / 27  \\
0.15~~--~0.20 & 0.174 & 0.062  & $\pm$ & 0.026  & $\pm$ & 0.010  &  18 / 27  \\
0.20~~--~0.30 & 0.247 & 0.058  & $\pm$ & 0.025  & $\pm$ & 0.009  &  25 / 27  \\
0.30~~--~0.50 & 0.390 & 0.025  & $\pm$ & 0.011  & $\pm$ & 0.004  &  48 / 27  \\
0.50~~--~0.70 & 0.590 & 0.0044 & $\pm$ & 0.0018 & $\pm$ & 0.0007 &  32 / 27  \\\hline
\end{tabular}
\caption[in]
        {
Differential $\lft$ distribution as function of the 
scaled $\lft$ energy $x_E^{\lft}$. The first error is the fit error, the 
second error the systematic error (see Section \ref{s:l1520rec}).
The $\chi^2/N_{df}$ specifies the quality of the fits.
\label{t:l1520dndxe}
}
\end{center}
\end{table}

The measured scaled energy $x_E^{\lft}$ distribution is given in 
Table~(\ref{t:l1520dndxe})
and shown in Figure~(\ref{f:l1520dndxe})
compared to the \opal result~\cite{opal_res_paper}. 
At low $x_E$ both measurements agree within error. 
For $x_E>0.3$ this measurement yields a rate about three times higher.
Note that in this energy range a clear $\lft$ signal is observed 
(see Figure ~(\ref{f:l1520mass}b)).
The measurements are compared to predictions of modified \herwig and \jsf models.
$\lft$ production has been implemented in these models either by 
replacing the $\Sigma^{*0}(1385)$
by the $\lft$ in the case of \jsf or by adding only the $\lft$ to the particle
list in case of \herwig. The predicted rates should not be expected
to be well reproduced by the models and the model predictions 
have been renormalised to the observed $\lft$ rate. 
The general shape of the fragmentation function is reproduced well
by both models.

\begin{figure}[tbh]
\wi 0.6\textwidth
\center{\epsfig{file=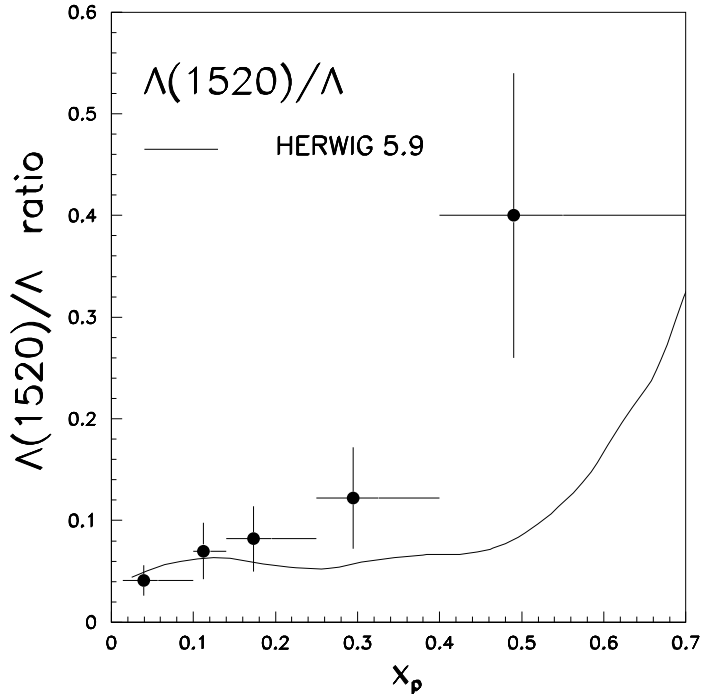,width=\wi}}
\caption[ratio l1520]
{
\label{f:l1520ratio}
Ratio of the differential $x_p$ distributions of $\lft$'s and $\Lambda$'s.
$\Lambda$ data are taken from  \cite{alephlambda}.
}
\end{figure}

In order to demonstrate the importance of $\lft$ production
in  Figure~(\ref{f:l1520ratio})
the ratio of $\lft$ to $\Lambda$ production is shown as function of the
scaled momentum $x_p$. For this comparison the
measurement of \cite{alephlambda} is taken as it covers a similar range in
scaled momentum like this $\lft$ measurement. It is seen that at small $x_p$,
$\lft$ production is about a factor 20 less than $\Lambda$ production.
At large $x_p$, however, this reduces to a factor $\sim$2.5. 
Such a behaviour would be expected from general fragmentation dynamics
due to the higher mass of the $\lft$.
An increase of this ratio is also expected if many $\Lambda$'s 
stem from resonance decays.
Finally it is interesting to note that the ratio of $\lft$ to proton production
is identical, within error, at low energies~\cite{argus} and in 
hadronic Z decays (as calculated from this result and \cite{EPJC5_585}).

\subsection{Discussion on baryon multiplicities}
\begin{figure}[tbh]
\center{\epsfig{width=0.9\textwidth,file=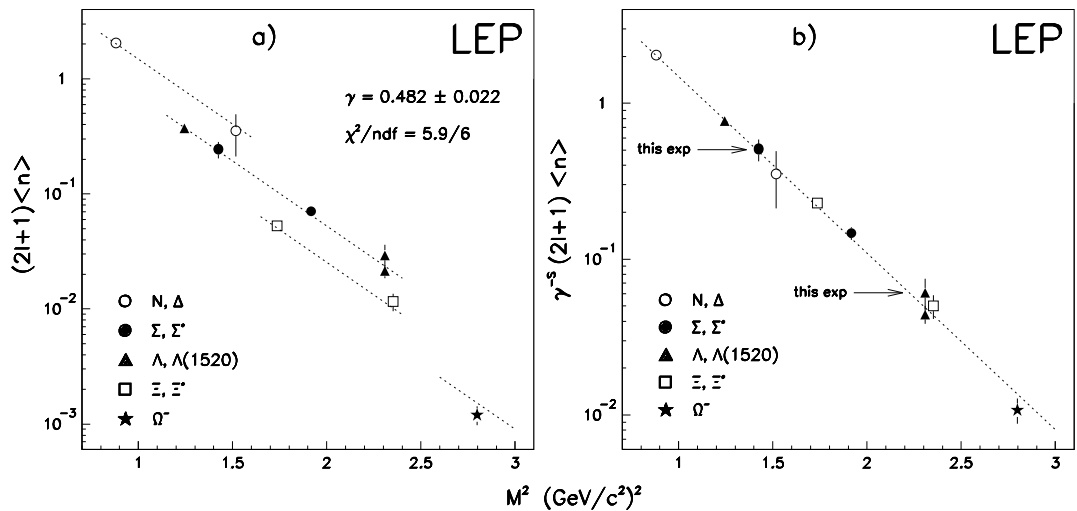}}
\caption[magicplots]
{
\label{f:bplot} 
a) Sum of the baryon production rates of all states of an isomultiplet
as a function of the squared baryon mass.
Data were taken from \cite{bary_multis}.
b) Sum of the baryon production rates of all states of an isomultiplet
weighted by $\gamma^{-S}$ as a function of the squared baryon mass.
}
\end{figure}

The measured $\sm$ and $\lft$ production rates may now be more generally compared
to the \lep average values for all baryons~\cite{bary_multis}.
In Figure~(\ref{f:bplot}a) the sum of the production rates 
of all states of 
an isomultiplet from the well known baryon octet and decuplet 
and for the orbitally excited isoscalar $\Lambda$(1520) are
shown as a function of
the corresponding particle mass squared, $M^2$.
In the case that not all states of an isomultiplet are measured at \lep, 
equal production rates for
these states are assumed. Therefore the vertical axis of the 
Figure~(\ref{f:bplot}a) is
denoted by $(2I+1)\,{\langle n \rangle}$,
where $\langle n \rangle$ is the mean number of a given particle per hadronic Z$^0$ decay.
It is seen that the mass dependence of the 
production rates is almost identical for the following sets of baryons: 
\begin{enumerate}
\item
$N$, $\Delta$ with strangeness S=0; 
\item
$\Sigma$, $\Sigma^*$, $\Lambda$ and $\Lambda$(1520) with S=1 and
\item
$\Xi$, $\Xi^*$ with S=2.
\end{enumerate}
Finally the $\Omega^-$ rate is well predicted if the same mass dependence
with an additional suppression for the higher strangeness (S=3) is assumed,
as taken from the difference of the first and second or second and third set,
respectively.
All data points from these three sets and the $\Omega^-$
are well fitted ($\chi^2/ndf=5.9/6$) by the ansatz (see dotted lines in
Figure~(\ref{f:bplot}a))
\begin{equation}
\gamma^{-S}\,(2I+1)\,{\langle n \rangle} = A\,\exp(-b\,M^2).
\label{e:magic}
\end{equation}
The values of the fitted parameters are:
$A=20.1 \pm 1.5$, $\gamma = 0.482 \pm 0.022$ and $b = 2.61 \pm 0.08$ 
(\gev/$c^2$)$^{-2}$. 
As decays of high mass particles feed down to lower mass states, it is to
be expected that especially the slope parameter $b$ for primary produced
baryons differs from the fitted value.
If the production rates are weighted by $\gamma^{-S}$ a universal
mass dependence is observed for all baryons (see Figure~(\ref{f:bplot}b)).

A similarly simple behaviour was found for scalar, vector and tensor mesons 
\cite{panic99}. 
Note that for mesons the production rates per spin and isospin projection  
were used. This is implicit in the figures shown in \cite{panic99}, as the
experimental rates for mesons are customarily specified for each isospin state
individually. 
The mass dependence for mesons, contrary to the baryonic case is exponential
in mass, $M$, thus for mesons
\begin{equation}
\gamma^{-k}\,\frac{1}{2J+1}\,{\langle n \rangle} = A\,\exp(-b\,M)~~~,
\end{equation}
where $k$ is the number of $s$ and $\bar{s}$ quarks in the meson.
For baryons and mesons an almost identical value of $\gamma \approx 0.5$ 
was found.
A stronger fall-off with the mass for baryons, compared to mesons, 
is to be expected as baryons are produced in pairs.

\section{Conclusions \label{s:conclusion}}

The differential cross-section of the $\sm$ hyperon has been measured in
multihadronic Z decays by reconstructing the kink between the
$\sm$ and the outgoing pion in the decay $\smdec$.
The measured production rate is
\begin{displaymath}
  \langle N_{\sm}/N_{\rm Z}^{\rm had}\rangle = 0.081 \pm 0.002 \mbox{ (stat.)}
                    \pm 0.006 \mbox{ (syst.)}
                    \pm 0.008 \mbox{ (extr.)}~~~.
\end{displaymath}
This result is about 20\% above the prediction of the \jsf 
model, but nevertheless compatible.
The shape of the differential cross-section is well described in this model.

The differential cross-section of the orbitally excited $\lft$ baryon has
been measured relying strongly on the particle identification capabilities of 
the \delphi detector.
The $\lft$ is the only measured orbitally excited baryon produced in 
fragmentation at \lep.
Its production rate is found to be
\begin{displaymath}
\langle N_{\lft} / N_{\rm Z}^{\rm had} \rangle = 
            0.0293 \pm 0.0049 {\rm (fit)} 
           \pm 0.0047  {\rm (syst.)} \pm 0.0003 {\rm (extr.)}~~~.
\end{displaymath}
This rate is similar to that of the $\Sigma^{*\pm}(1385)$ and suggests
that also other orbitally excited baryonic states are produced in fragmentation.
The shape of the $\lft$ fragmentation function is well described by the
\jsf and \herwig fragmentation model if $\lft$ production is introduced.

It has also been shown that the production rates of all baryonic states measured so far at
\lep can be parametrised by a phenomenological law.

\subsection*{Acknowledgements}
\vskip 3 mm
 We are greatly indebted to our technical 
collaborators, to the members of the CERN-SL Division for the excellent 
performance of the LEP collider, and to the funding agencies for their
support in building and operating the DELPHI detector.\\
We acknowledge in particular the support of \\
Austrian Federal Ministry of Science and Traffics, GZ 616.364/2-III/2a/98, \\
FNRS--FWO, Belgium,  \\
FINEP, CNPq, CAPES, FUJB and FAPERJ, Brazil, \\
Czech Ministry of Industry and Trade, GA CR 202/96/0450 and GA AVCR A1010521,\\
Danish Natural Research Council, \\
Commission of the European Communities (DG XII), \\
Direction des Sciences de la Mati$\grave{\mbox{\rm e}}$re, CEA, France, \\
Bundesministerium f$\ddot{\mbox{\rm u}}$r Bildung, Wissenschaft, Forschung 
und Technologie, Germany,\\
General Secretariat for Research and Technology, Greece, \\
National Science Foundation (NWO) and Foundation for Research on Matter (FOM),
The Netherlands, \\
Norwegian Research Council,  \\
State Committee for Scientific Research, Poland, 2P03B06015, 2P03B1116 and
SPUB/P03/178/98, \\
JNICT--Junta Nacional de Investiga\c{c}\~{a}o Cient\'{\i}fica 
e Tecnol$\acute{\mbox{\rm o}}$gica, Portugal, \\
Vedecka grantova agentura MS SR, Slovakia, Nr. 95/5195/134, \\
Ministry of Science and Technology of the Republic of Slovenia, \\
CICYT, Spain, AEN96--1661 and AEN96-1681,  \\
The Swedish Natural Science Research Council,      \\
Particle Physics and Astronomy Research Council, UK, \\
Department of Energy, USA, DE--FG02--94ER40817. \\

\end{document}